\documentclass[aps,prx,groupedaddress,reprint,showpacs,pdftex]{revtex4-1}
\usepackage{graphicx}
\usepackage{tabularx}
\usepackage{multirow}
\usepackage{longtable}
\usepackage{dcolumn}
\usepackage{bm}
\usepackage{amsmath}
\usepackage{amssymb}
\usepackage{txfonts}
\usepackage{url}
\usepackage{ulem}
\usepackage[usenames,dvipsnames]{color}
\definecolor{Gray}{gray}{0.0}
\definecolor{lightGray}{gray}{0.35}

\begin{document}
\title{
High-$T_c$ superconducting hydrides 
formed by LaH$_{24}$ and YH$_{24}$ cage structures as basic blocks.
}
\author{
  Peng Song$^{1}$,
  Zhufeng Hou$^{2}$,
  Pedro Baptista de Castro$^{3,4}$,
  Kousuke Nakano$^{1,5}$,
  Kenta Hongo$^{6}$,
  Yoshihiko Takano$^{3,4}$,
  Ryo Maezono$^{1}$ \\}

\affiliation{\\
  $^1$School of Information Science, JAIST,
  Asahidai 1-1, Nomi, Ishikawa 923-1292, Japan\\
  \\
  $^2$State Key Laboratory of Structural Chemistry,
  Fujian Institute of Research on the Structure of Matter,
  Chinese Academy of Sciences, Fuzhou 350002, China \\
  \\
  $^3$National Institute for Materials Science, 1-2-1 Sengen, Tsukuba, Ibaraki 305-0047, Japan\\
  \\
  $^4$University of Tsukuba, 1-1-1 Tennodai, Tsukuba, Ibaraki 305-8577, Japan\\
  \\
  $^5$International School for Advanced Studies (SISSA),
      Via Bonomea 265, 34136, Trieste, Italy\\
  \\
  $^6$Research Center for Advanced Computing
      Infrastructure, JAIST, Asahidai 1-1, Nomi,
      Ishikawa 923-1292, Japan\\
  \\
}

\vspace{10mm}

\date{\today}
\begin{abstract}
Based on recent studies regarding high-temperature (high-$T_c$) La-Y ternary hydrides
(e.g., $P{\bar{1}}$-La$_2$YH$_{12}$, $Pm{\bar{3}}m$-LaYH$_{12}$,
and $Pm{\bar{3}}m$-(La,Y)H$_{10}$ with a maximum $T_c \sim 253$ K), 
we examined the phase and structural stabilities of the 
 (LaH$_6$)(YH$_6$)$_y$ series as 
high-$T_c$ ternary hydride compositions 
using a genetic algorithm and {\it ab initio} 
calculations. 
Our evaluation showed that 
the $Pm\bar{3}m$-LaYH$_{12}$ reported in the previous study 
was unstable during decomposition into
$R\bar{3}c$-LaH$_{6}$ + $Im\bar{3}m$-YH$_{6}$. 
We also discovered new crystal structures, 
namely $Cmmm$-LaYH$_{12}$ ($y=1$), 
$R\bar{3}c$-LaYH$_{12}$ ($y=1$), $Cmmm$-LaY$_3$H$_{24}$ ($y=3$), 
and $R\bar{3}$-LaY$_3$H$_{24}$ ($y=3$), 
showing stability against such decomposition. 
While $R\bar{3}c$ ($y=1$) and $R\bar{3}$ ($y=3$) 
did not exhibit superconductivity 
owing to the extremely low density of states at the Fermi level, 
$Cmmm$ phases exhibited a 
$T_{c}$ of approximately 140~K at around 200~GPa 
owing to the extremely high electron--phonon coupling 
constant ($\lambda$ = 1.876 for LaYH$_{12}$). 
By the twice longer stacking for $Cmmm$-LaY$_3$H$_{24}$, 
the coupling constant increased owing to the 
chemical pressure of Y, leading to 
a slightly increased $T_{c}$.
\end{abstract}
\maketitle

 \section{Introduction}
 \label{sec.intro}
Metal hydrides have always been considered ideal
candidates for high-temperature (high-$T_c$)
superconductors owing to their ultrahigh phonon
vibration frequency.~\cite{2004ASH,1968ASH}
However, realizing superconductivity
by keeping them under the metallicity gradient
often requires extremely high pressures.~\cite{2004ASH,1968ASH,2008ERE,2015DRO,2015SZC,2021SNI}
Therefore, theoretical simulations play an extremely
important role in search for novel hydride superconductors.~\cite{2006GLA,2012WAN,2011LON}
Through theoretical simulations, almost all the predictions
for binary hydrides have been completed, and studies have successfully predicted the room-temperature
superconductivity of YH$_{10}$.~\cite{2017PEN,2019HEI}
The successful prediction of room-temperature superconductivity in
ternary Li$_{2}$MgH$_{16}$~\cite{2019SUN}
and the experimental discovery
of room-temperature superconductivity in the C--S--H system~\cite{2020SNI}
have pushed research on
superconducting ternary hydrides system to a climax.
\begin{figure*}
  \begin{center}
    \includegraphics[width=0.7\linewidth]{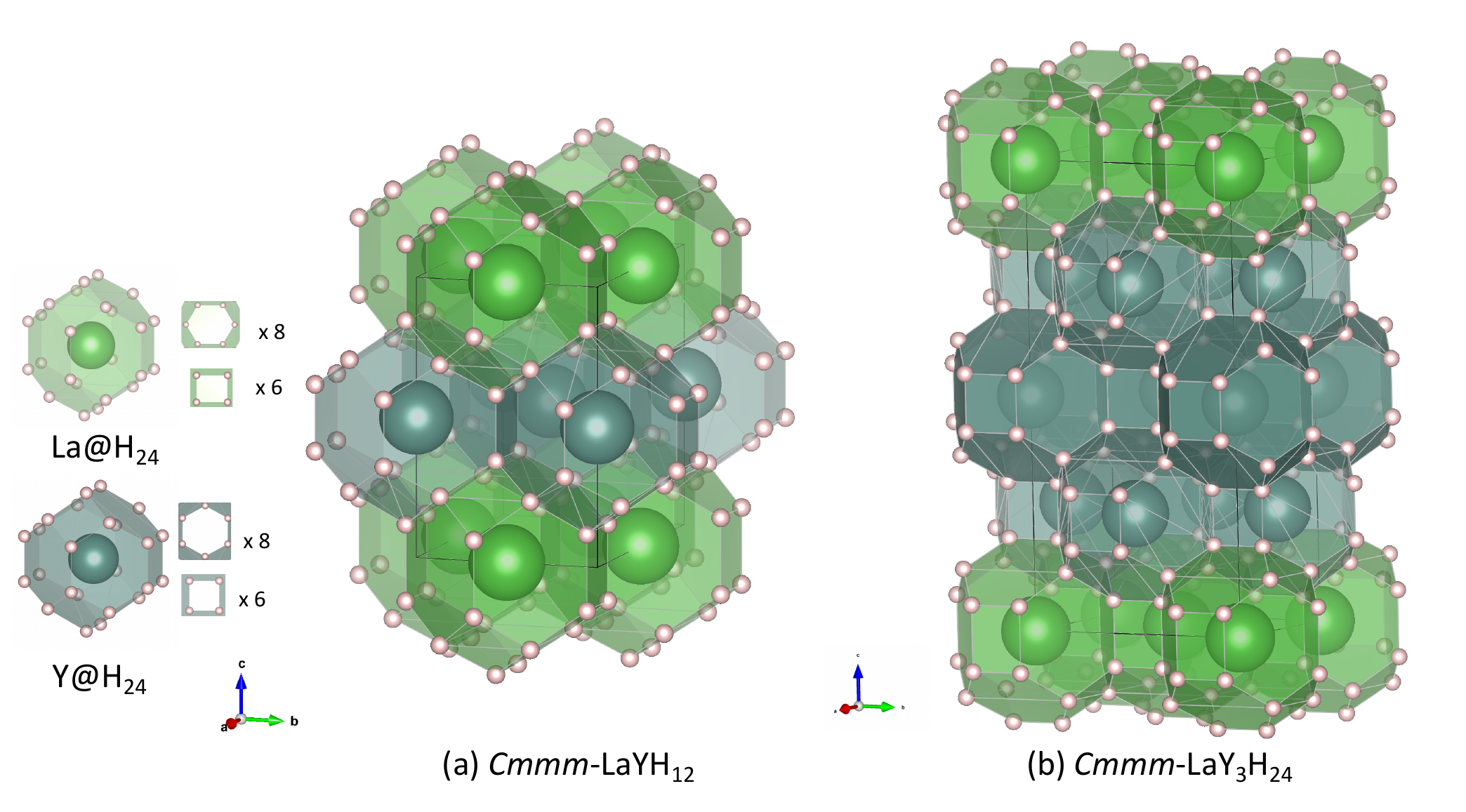}
    \caption{
      Clathrate structure of LaYH$_{12}$ (left panel, at 200~GPa)
      and LaY$_{3}$H$_{24}$ (right panel, at 180~GPa)
      with $Cmmm$, 
      consisting of La-centered H$_{24}$ cages
      and Y-centered H$_{24}$ cages. 
Each H$_{24}$ cage is surrounded by 
six squares and eight hexagons.
    }
    \label{fig.structure}
  \end{center}
\end{figure*}

\vspace{2mm}
However, owing to the diversity of ternary hydrides,
the search for new high-$T_c$ superconductors appears 
very slow, and only $10 \sim 20$ potential superconducting hydrides
have been identified.~\cite{2019SUN,2020SNI,2017MA_b,2017MA,2018LI,2019LIA_b,
2017RAH,2019LIA,2020WEI,2019XIE,2017KOK,2019SHA,2019SHA,
2020ZHA_a,2020ZHA_b,2020DIC,2020GUO,2020CUI,2020LV,2020YAN,2015MUR,2019MEN}
From the information obtained regarding binary 
superconductors, the crystal structure significantly influences
the $T_c$.
For example, some ultrahigh-$T_c$ superconductors
often appear in cubic systems such as
($Fm\bar{3}m$-YH$_{10}$, $Fm\bar{3}m$-LaH$_{10}$, and
$Im\bar{3}m$-H$_{3}$S, among others. 
~\cite{2017PEN,2019SOM,2015ERR}
However, in ternary hydrides, 
the stability of the cubic structure is very poor, and many ternary systems
may exhibit phase decomposition relative to binary hydrides.
For example, CSH$_{7}$,~\cite{2020CUI}
is stable relative to C, S, and H,  
but not to binary hydrides CH$_{4}$~\cite{2010GAO} and H$_{3}$S~\cite{2015ERR},
possibly leading to phase decomposition. 
Owing to phase decomposition,
it is highly unlikely  that
these compounds can be synthesized experimentally. 
Therefore, it is extremely important to ensure
their stability relative to 
the constituent elements and
binary materials for  predicting
ternary hydrides. 

\vspace{2mm}
For searching ternary hydrides, 
there are two major approaches aiming at obtaining a high $T_c$, 
one focusing on SH$_x$ doped with La, Se, Te, and Cl {\it etc.},
~\cite{2019LIA,2018LIU,2020CHA,2018NAK} 
and the other on La- or Y-hydrides 
doped with metallic elements such as 
CaYH$_{12}$, CaMgH$_{12}$, KScH$_{12}$, and LaKH$_{12}$.
~\cite{2019LIA,2020SUK,2019ISH,2021SON}
With no significant improvement in $T_c$ observed so far 
in the latter approach, the required pressure to realize 
superconductivity in these materials is considerably reduced, 
thereby providing the possibility of their experimental synthesis.
Experiments involving the La--Y--H system 
report high-$T_c$ superconductivity
in (La,Y)H$_{10}$.~\cite{2020SEM}

\vspace{2mm}
Based on these experiments, theoretical 
predictions of superconductivity in 
LaYH$_{12}$, La$_{2}$YH$_{18}$, La$_{4}$YH$_{30}$
{\it etc.} 
have been also reported.~\cite{2020SEM}
These compositions are referred to as composites 
of (LaH$_{6}$)$_x$ + YH$_{6}$. 
The symmetry of their predicted structures is
considerably low mainly being $P\bar{1}$ with 
highest mostly $Pm\bar{3}m$ for LaYH$_{12}$. 
Examining its decomposition enthalpy 
revealed that the structure may exhibit phase 
decomposition under high pressure 
toward $R\bar{3}c$-LaH$_{6}$ + $Im\bar{3}m$-YH$_{6}$, 
being unlikely to be synthesized experimentally.~\cite{2019HEI,2017PEN}
We then determined whether LaYH$_{12}$ has 
other structures stable against decomposition based
on the USPEX fixed composition method. ~\cite{2006GLA}
Composition LaYH$_{12}$ 
was also found in our 
machine learning search based on a
gradient boosting tree~\cite{2021SON}, and it exhibited a high $T_c$. 
When searching more general forms, namely
(LaH$_6$)$_x$(YH$_6$)$_y$, 
we restricted the range by fixing $x=1$ 
because the preceding work has reported 
instability along the $x$-axis. 
Along the $y$-axis, we found 
$Cmmm$-LaYH$_{12}$ ($y=1$) 
and $Cmmm$-LaY$_3$H$_{24}$ ($y=3$) 
exhibiting a high $T_c \sim$ of 140~K. 
These compounds are formed
through stacking of 
LaH$_{24}$ and YH$_{24}$ cages, 
as shown in Fig.~\ref{fig.structure}. 
A slightly higher $T_c$ for 
$Cmmm$-LaY$_3$H$_{24}$ compared to that for
LaYH$_{12}$ can be realized 
by enhancing the chemical pressure for Y 
through further stacking. 

 \section{Results}
\begin{figure*}
  \begin{center}
    \includegraphics[width=\linewidth]{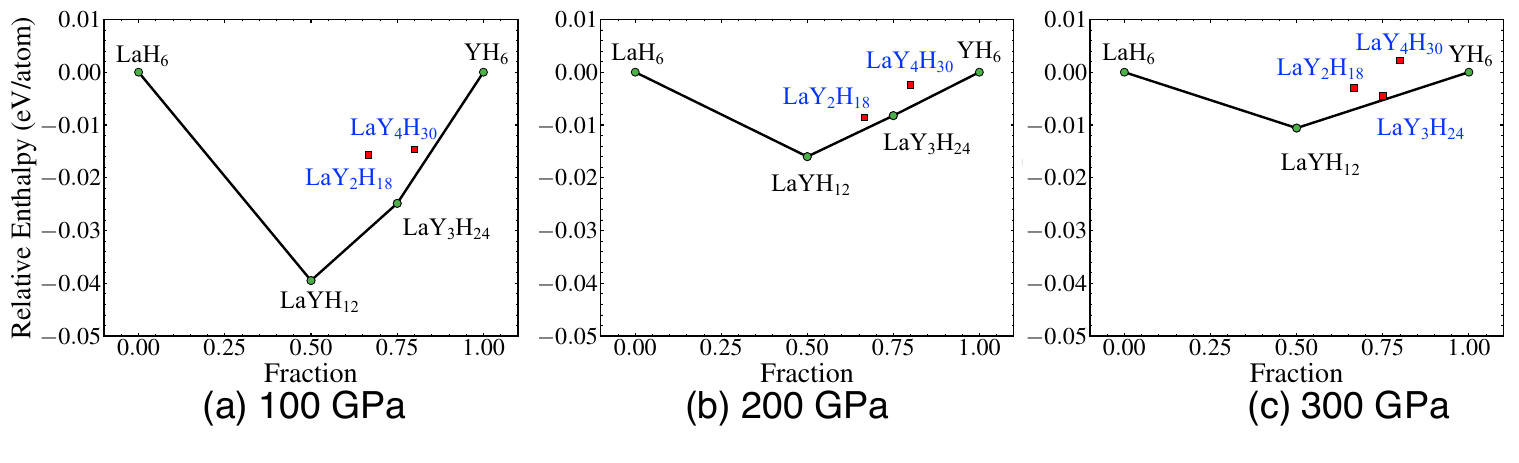}
    \caption{
     Convex hull of La--Y--H at pressures of 100~GPa, 200~GPa, and 300~GPa.
    }
    \label{fig.convex_hull}
  \end{center}
\end{figure*}

\begin{figure*}
  \begin{center}
    \includegraphics[width=0.8\linewidth]{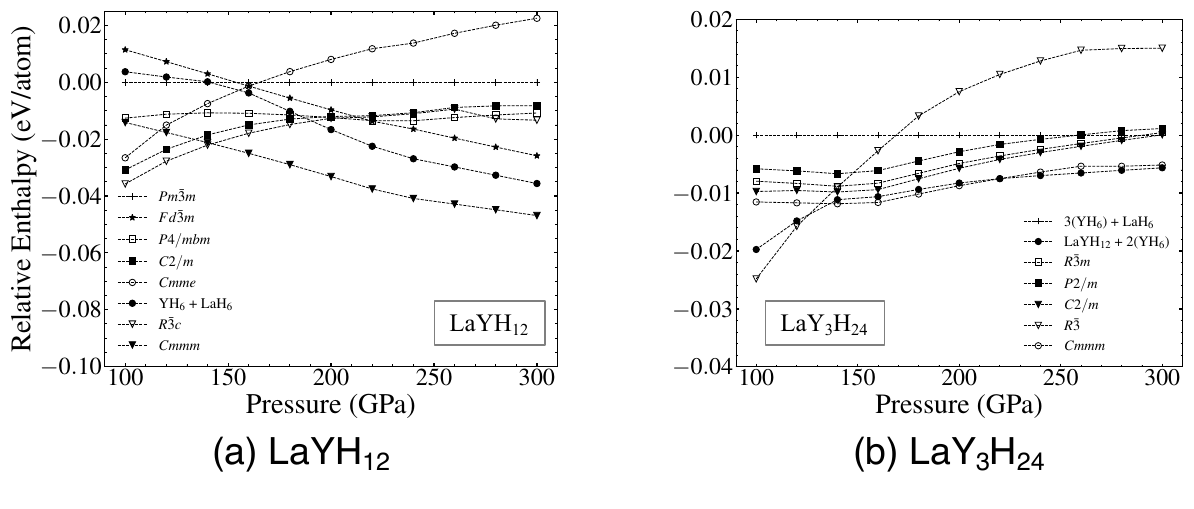}
    \caption{
        Comparisons of candidate structure enthalpies for LaYH$_{12}$ [panel (a)] and 
        LaY$_{3}$H$_{24}$ [panel (b)]. 
        Such structures giving higher values than those of 
        'YH$_{6}$ + LaH$_{6}$' (for the left panel),
        '3(YH$_{6}$) + LaH$_{6}$' (for the right), and 
        'LaYH$_{12}$ + 2(YH$_{6}$)' (for the right), 
        are predicted 
        to exhibit instability toward decomposition 
        into these binary compounds. 
    }
    \label{fig.enthalpy}
  \end{center}
\end{figure*}

Fig.~\ref{fig.convex_hull} shows 
the convex hull of our search 
for LaH$_{6}$(YH$_{6}$)$_x$ ($x$ = 1-4) structures
at 100, 200, and 300 GPa. 
LaYH$_{12}$ ($x$=1) and 
LaY$_{3}$H$_{24}$ ($x$=3) are stable phases, whereas other 
LaY$_{2}$H$_{18}$ and LaY$_{4}$H$_{30}$ structures
are likely to be decomposed at the abovementioned pressures . 
For the stable compositions, 
namely LaYH$_{12}$ and LaY$_{3}$H$_{24}$, 
we compared their relative enthalpies 
with those of the candidate structures over the pressure range, 
as shown in Fig.~\ref{fig.enthalpy}. 
The plot for LaYH$_{12}$ indicates 
that the $Pm\bar{3}m$ structure proposed 
recently~\cite{2020SEM} 
exhibits instability toward 
decomposition into 
binary compounds, namely
YH$_{6}$($Im\bar{3}m$) and 
LaH$_{6}$($R\bar{3}c$). 

\vspace{2mm}
For the entire pressure range, 
more stable structures are predicted 
as $R\bar{3}c\to Cmmm$ with 
the transition occurring at approximately $P = 140$~GPa, 
without instability toward decomposition. 
For LaY$_{3}$H$_{24}$, stable structures form
as $R\bar{3} \to Cmmm$, preventing its 
decomposition either to LaYH$_{12}$ + 2(YH$_{6}$) 
or 3(YH$_{6}$) + LaH$_{6}$. 
As explained later, $R\bar{3}c$-LaYH$_{12}$ 
and $R\bar{3}$-LaY$_{3}$H$_{24}$ 
exhibit an extremely low density of states (DOS) at the 
Fermi level, $D\left(\varepsilon_F \right)$,
leading to a considerably low $T_c$. 
In contrast, $Cmmm$ compounds are exhibit a
higher $T_c$, and their structures 
are shown in Fig.~\ref{fig.structure}. 
Detailed geometries for each structure 
are provided in Table~\ref{table.crytal_struc} 
in the Supplemental Information (S.I.).

\vspace{2mm}
As observed in the Fig. 1, 
the structures are mostly 
clathrate structures formed
by units with 
La@H$_{24}$ and Y@H$_{24}$ cage structures. 
Several other clathrate superconductors 
have been identified,
~\cite{2017PEN,2012WAN_b,2021WAN,2019SUN,2019HEI,2019SAL,2019XIE} 
and hence it would be plausible 
for this structure to exhibit superconductivity. 
Different periodicities 
of the La- and Y-layers along the $c$-direction 
correspond to LaYH$_{12}$ 
[(La/Y)(La/Y)$\cdots$] 
and LaY$_{3}$H$_{24}$ 
[(La/Y/Y/Y)(La/Y/Y/Y)$\cdots$], respectively. 

\begin{figure}
  \begin{center}
    \includegraphics[width=\linewidth]{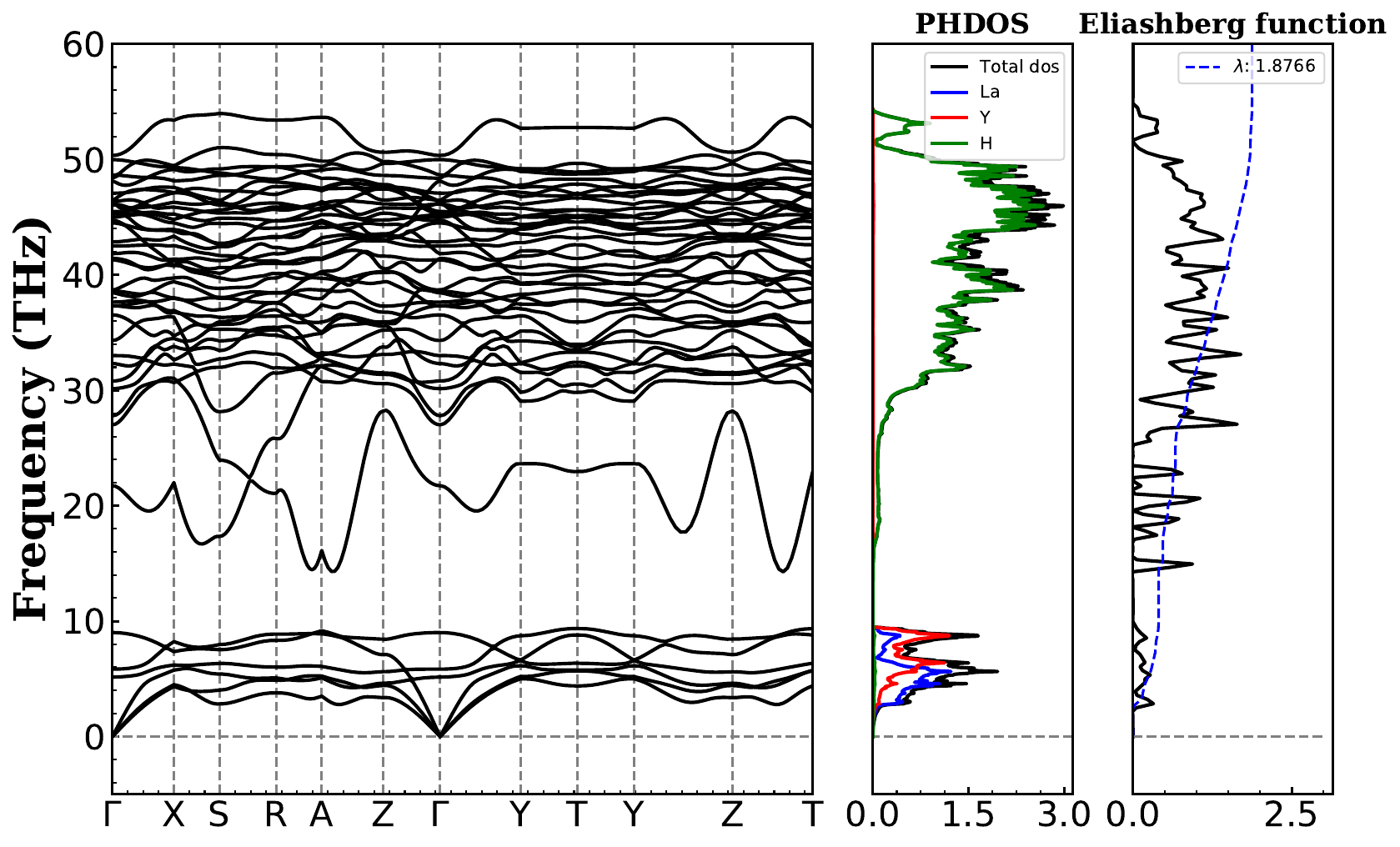}
    \caption{
      Phonon dispersions, projected phonon
      densities of states, and Eliashberg spectral function 
      for $Cmmm$-LaYH$_{12}$ clathrate at 200~GPa.
    }
    \label{fig.repPhonon}
  \end{center}
\end{figure}

\vspace{2mm}
Fig.~\ref{fig.repPhonon} shows 
the phonon dispersions evaluated for 
the clathrate $Cmmm$ LaYH$_{12}$. 
We confirmed no imaginary modes, 
thus confirming its structural stability. 
The phonon dispersions for other structures are 
given in the S.I., 
confirming that all the candidates 
predicted from Fig.~\ref{fig.enthalpy}, namely
$Cmmm$-LaYH$_{12}$, $Cmmm$-LaY$_{3}$H$_{24}$
$R\bar{3}c$-LaYH$_{12}$, and 
$R\bar{3}$-LaY$_{3}$H$_{24}$, 
have no imaginary modes. 
In contrast, $Pm\bar{3}m$-LaYH$_{12}$ 
proposed recently~\cite{2020SEM} 
exhibits imaginary modes, 
as shown in Fig.~\ref{fig.phonon_LaYH12_221}, 
consistent with the observations from Fig.~\ref{fig.enthalpy} 
in terms of the instability. 

\vspace{2mm}
Applying the Allen--Dynes formalism,
~\cite{2009GIA,2017NAK,2016NAK} 
we estimated $T_c$ as summarized in 
Table~\ref{table.Pdep}. 
For LaYH$_{12}$, 
$T_c$ is in the $130.6 \sim 140.55$~[K] range,
depending on the choice of parameter 
$\mu = 0.10 \sim 0.13$ for the Coulomb 
interaction. 
A slightly higher $T_c$ realized 
for LaY$_{3}$H$_{24}$ is attributed 
to the higher $D\left(\varepsilon_F\right)$ 
due to the compressed chemical pressure 
for the Y-site with the (La/Y/Y/Y) stacking structure, 
compared to the (La/Y)(La/Y) periodicity.
\begin{table*}
 \begin{center}
   \caption{
     $T_c$ (expressed in [K]) estimated using the Allen--Dynes formula through
     {\it ab initio} phonon calculations
     for LaYH$_{12}$ and LaY$_{3}$H$_{24}$
     at each pressure.
     $\lambda$ and $\omega_{\rm log}$ (expressed in [K]) are
     the parameters appearing in the formula.
    The values of $D\left(\varepsilon_F\right)$ are
     are expressed in [states/eV/atom].
     \label{table.Pdep}
   }
\begin{tabular}{rccrll}
\hline
& $P$~[GPa] & $\lambda$ & $\omega_{\rm log}$
& $D\left(\varepsilon_F\right)$ & $T_c$ ($\mu$ = 0.1 - 0.13) \\
\hline
LaYH$_{12}$& 200         & 1.876    & 1022.54   & 0.097 &140.55 - 130.61     \\
LaYH$_{12}$& 250         & 1.618    & 1051.56   & 0.093 &128.42 - 117.96     \\
LaY$_{3}$H$_{24}$& 180   & 2.452     & 891.49   & 0.100 &145.31 - 137.11     \\
\hline
\end{tabular}
 \end{center}
\end{table*}

 \section{Discussions}
As observed for 
$Fd\bar{3}m$-Li$_{2}$MgH$_{16}$, a room-temperature superconductor
~\cite{2019SUN}, 
the key factors for achieving a high $T_c$ 
are its high DOS and high phonon frequency,
which can be measured by $\lambda$ and $\omega_{\rm log}$, respectively.
%
%
$\lambda = 2$ would be the typical criterion for 
realizing superconductivity 
($\lambda$ = 3.35 for $Fd\bar{3}m$-Li$_{2}$MgH$_{16}$).
~\cite{2019SUN} 

\vspace{2mm}
In the S.I., 
we have provided the band structures and 
Fermi surfaces of $R\bar{3}c$-LaYH$_{12}$ 
and $R\bar{3}c$-LaYH$_{12}$
(Figs.~\ref{fig.LaYH12_band_dos_167} 
and \ref{fig.LaY3H24_band_dos_148}). 
The Fermi surfaces and the 
DOS values on these surfaces are extremely small and are insufficient 
to achieve superconductivity. 

\vspace{2mm}
In contrast, for the $Cmmm$ structure, 
sufficiently large Fermi surfaces and high DOS 
values lead to higher $\lambda$, 
thus resulting in  superconductivity. 
Examining the phonon dispersions and 
Eliashberg spectral functions 
(Figs.~\ref{fig.repPhonon}, \ref{fig.phonon_LaYH12_65_250GPa}
and \ref{fig.phonon_LaY3H24_65}) reveals that 
higher $\lambda$ values can be attributed to 
the contribution from the phonon branches 
appearing at higher ranges ($>$ 10~THz), 
amounting to 96.2\%. 
These branches are formed by the vibration of 
hydrogen atoms, whereas 
those at lower ranges ($<$ 10~THz) 
by La and Y, negligibly contributing to $\lambda$. 

\vspace{2mm}
Applying higher pressures (200$\to$250~GPa) 
to $Cmmm$-LaY$_{3}$H$_{24}$ increases 
its phonon frequencies 
for the branches at higher ranges ~($>$ 10~THz), 
as observed in 
Figs.~\ref{fig.repPhonon} and 
\ref{fig.phonon_LaYH12_65_250GPa}. 
This trend is reflected as 
the increased $\omega_{\rm log}$ 
(Table~\ref{table.Pdep}), 
contributing toward increasing the $T_c$. 
However, this trend disappears 
when the DOS (around 4.2\%) decreases owing to 
the applied pressure, leading to a 
reduced $\lambda$ value. 
This decrease is more dominant  than
the increase in $\omega_{\rm log}$, 
and hence the applied pressure eventually 
lowers the $T_{c}$, as observed in Table~\ref{table.Pdep}.


\vspace{2mm}
The slightly higher $T_c$ determined for LaY$_{3}$H$_{24}$ 
is attributed to the enhanced $\lambda$, 
as observed in Table~\ref{table.Pdep}. 
In this case too, changes in $\omega_{\rm log}$ 
and $\lambda$ exhibit a cancelling relationship, 
causing only a slight increase in the $T_{c}$. 
Applying further pressure would mean increasing 
$\lambda$ further for achieving a higher $T_{c}$; 
however, according to the convex hull in our analysis, 
$Cmmm$-LaY$_{3}$H$_{24}$ 
exhibits instability toward 
phase decomposition above 220 GPa.

\begin{figure}
  \begin{center}
    \includegraphics[width=\linewidth]{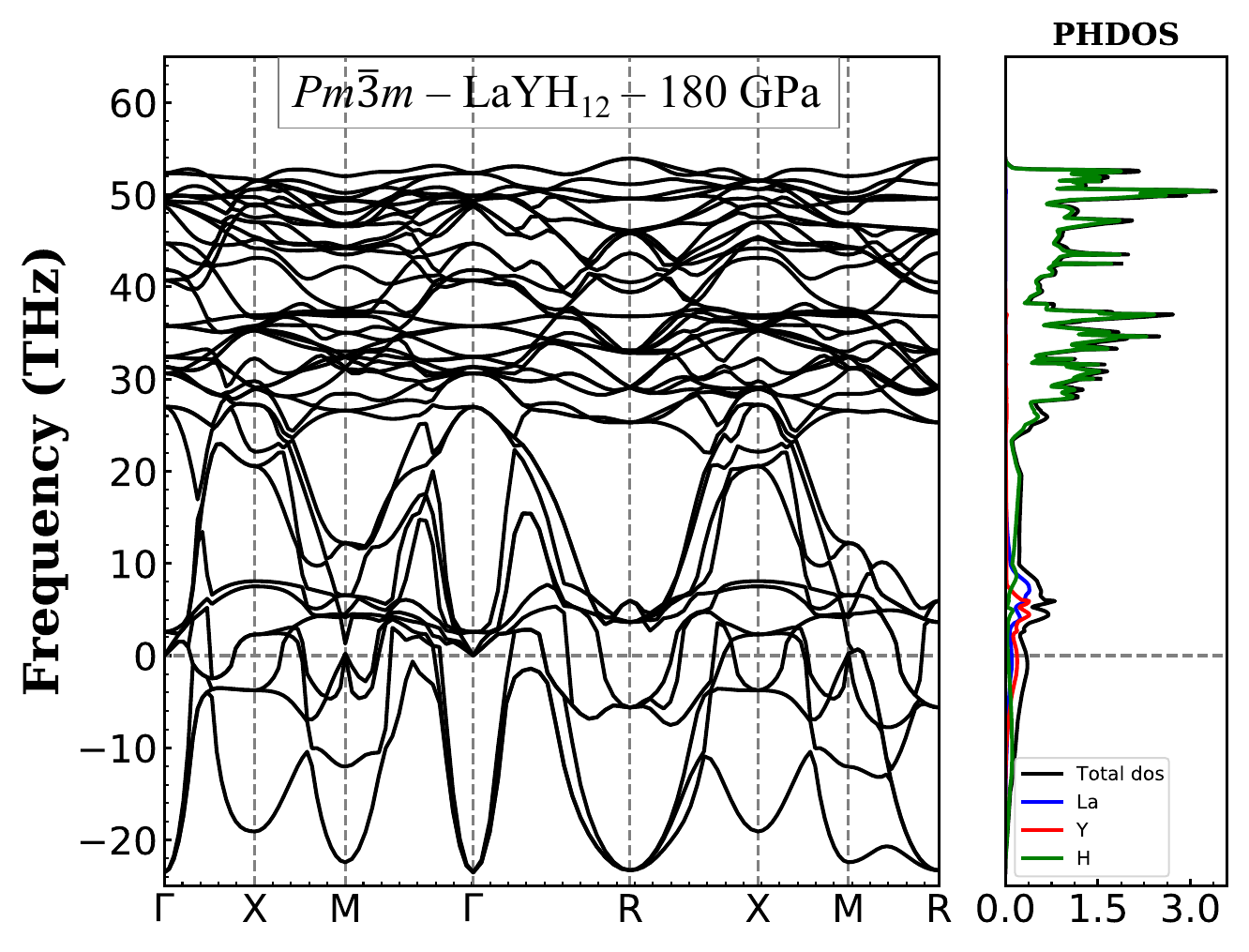}
    \caption{
      Phonon dispersions and projected phonon
      densities of states 
      for $Pm\bar{3}m$-LaYH$_{12}$ at 180~GPa. 
    }
    \label{fig.phonon_LaYH12_221}
  \end{center}
\end{figure}

\vspace{2mm}
$Pm\bar{3}m$-LaYH$_{12}$ 
exhibits a higher $T_c$~(203 K at 180~GPa)
~\cite{2020SEM}
However, for this structure, 
the imaginary phonon modes appeared 
as shown in Fig.~\ref{fig.phonon_LaYH12_221}, 
implying structural instability 
at least within the extent of the harmonic 
approximation. 
As shown in Fig.~\ref{fig.enthalpy}, 
another structure, $Fd\bar{3}m$-LaYH$_{12}$, 
is predicted to be more stable 
than $Pm\bar{3}m$. 
However, for both structures, 
the imaginary modes are found, without 
vanishing even on applying 
further pressure. 

\begin{figure*}
  \begin{center}
    \includegraphics[width=0.7\linewidth]{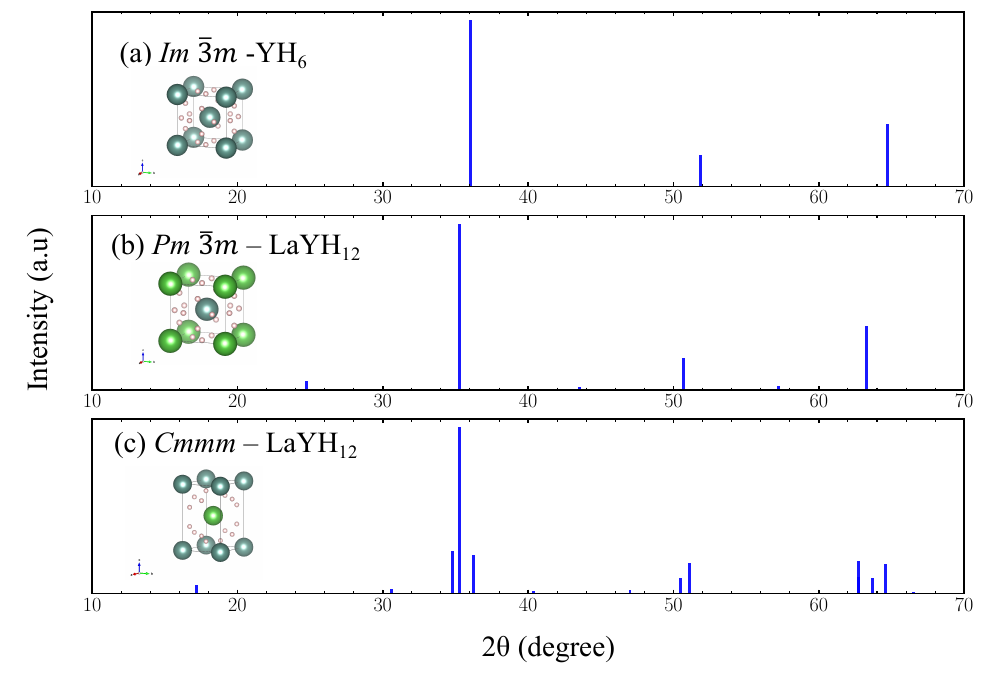}
    \caption{Structures and the 
      XRD patterns of (a) $Im\bar{3}m$-YH$_{6}$, (b) $Pm\bar{3}m$-LaYH$_{12}$, and 
      (c) $Cmmm$-LaYH12$_{12}$.
    }
    \label{fig.XRD}
  \end{center}
\end{figure*}

\vspace{2mm}
$Pm\bar{3}m$-LaYH$_{12}$ 
can be regarded as a 
compound based on 
the $Im\bar{3}m$-YH$_{6}$ dimer 
with La substituting for one Y 
(Fig.~\ref{fig.XRD}). 
The $Im\bar{3}m$-YH$_{6}$ 
actually appears as a 
stable structure in the 
convex-hull analysis given in 
Fig.~\ref{fig.convex_hull}. 
Our prediction, $Cmmm$-LaYH$_{12}$, 
is actually a distorted 
structure of $Pm\bar{3}m$-LaYH$_{12}$, 
showing considerably similar 
X-ray diffraction (XRD) peak patterns 
as well as that of $Fd\bar{3}m$, 
as shown in Fig.~\ref{fig.convex_hull}. 
Furthermore, the possibility of the 
structural transition 
between $Cmmm$ and $Pm\bar{3}m$ at 
higher temperatures is pointed out.
~\cite{2019SUN}

 \section{Conclusion}
 \label{sec.conc}
In summary, we investigated the possible crystal structures for 
(LaH$_6$)(YH$_6$)$_y$ compounds as candidates for forming ternary hydrides 
to achieve a high $T_c$ using a
generic algorithm implemented in USPEX. 
We determined the thermal stabilities
at $y=1$~(LaYH$_{12}$) and 
$y=3$~(LaY$_3$H$_{24}$). 
A recently reported structure, $Pm\bar{3}m$, 
appeared unstable toward
the decomposition, i.e.,
LaYH$_{12}$ ($Pm\bar{3}m$)
$\to$ LaH$_{6}$($R\bar{3}c$) + YH$_{6}$($Im\bar{3}m$). 
However, we discovered other new stable structures 
that did not exhibit decomposition, namely
$Cmmm$-LaYH$_{12}$, 
$R\bar{3}c$-LaYH$_{12}$, $Cmmm$-LaY$_3$H$_{24}$, 
and $R\bar{3}$-LaY$_3$H$_{24}$. 
Among these, the compounds with 
$Cmmm$ clathrate structures 
exhibited a $T_c\sim$ 130~[K] at approximately $P$ = 200~GPa, 
as estimated using the Allen--Dynes formalism. 

\section{Methods}
For the structural predictions, 
we used the USPEX code~\cite{2006GLA} combined with the {\it ab initio} kernel
in the Vienna ab initio simulation package 
(VASP).
~\cite{1993KRE,1994KRE,1996KRE_a,1996KRE_b}
For example, the LaYH$_{12}$ composition 
randomly generated 400 structures
ranging from monomers to tetramers with a
LaYH$_{12}$ unit as the ``initial generation.'' 
Each generation evolved 100 structures
with 40\% heredity, 40\% randomness,
10\% softmutation, and 10\% transmutation.
When no further evolution occurred for more than 
10 generations, the final structure 
was identified as the candidate crystal structure of the composition. 
For the identified candidates, we performed 
{\it ab initio} geometrical optimizations
using the Perdew--Burke--Ernzerhof (GGA--PBE)
functional for the exchange-correlation
functional.~\cite{1996PER}

\vspace{2mm}
For the most stable candidates
[$R\bar{3}c$~(100-140~GPa) and $Cmmm$~(140-300~GPa)
of LaYH$_{12}$, for example], 
we performed {\it ab intio} phonon 
calculations to evaluate their structural stabilities 
and $T_c$ using the Allen--Dynes formalism,~\cite{2009GIA,2017NAK,2016NAK} 
implemented in Quantum Espresso.~\cite{2009GIA} 
Computational conditions such as 
the cutoff energy of the plane wave basis set, 
sizes of the grid-mesh over the Brillouin zone 
{\it etc.} were determined 
such that
the energies sufficiently converged in 
their dependences. 
We finally used an $8\times 8\times 5$ $k$-mesh used in the self-consistent field 
convergence assisted by the 
Marzari--Vanderbilt smearing scheme.~\cite{1999MAR} 
The final energy values 
were estimated
through extrapolation of the smearing parameter toward zero. 
Mesh sizes for the phonon calculations 
were $16\times 16\times 10$ for the $k$- 
and $4\times 4\times 2$ for the $q$-mesh. 

 \section{Acknowledgments}
The computations in this work have been performed
using the facilities of
Research Center for Advanced Computing
Infrastructure (RCACI) at JAIST.
K.H. is grateful for financial support from
the HPCI System Research Project (Project ID: hp190169) and
MEXT-KAKENHI (JP16H06439, JP17K17762, JP19K05029, and JP19H05169).
R.M. is grateful for financial supports from
MEXT-KAKENHI (21K03400 and 19H04692), 
and the Air Force Office of Scientific Research
(AFOSR-AOARD/FA2386-17-1-4049;FA2386-19-1-4015)

\bibliographystyle{apsrev4-1}
\bibliography{references}

\clearpage
\section{Supplemental Information}
\begin{figure*}
  \begin{center}
    \includegraphics[width=0.5\linewidth]{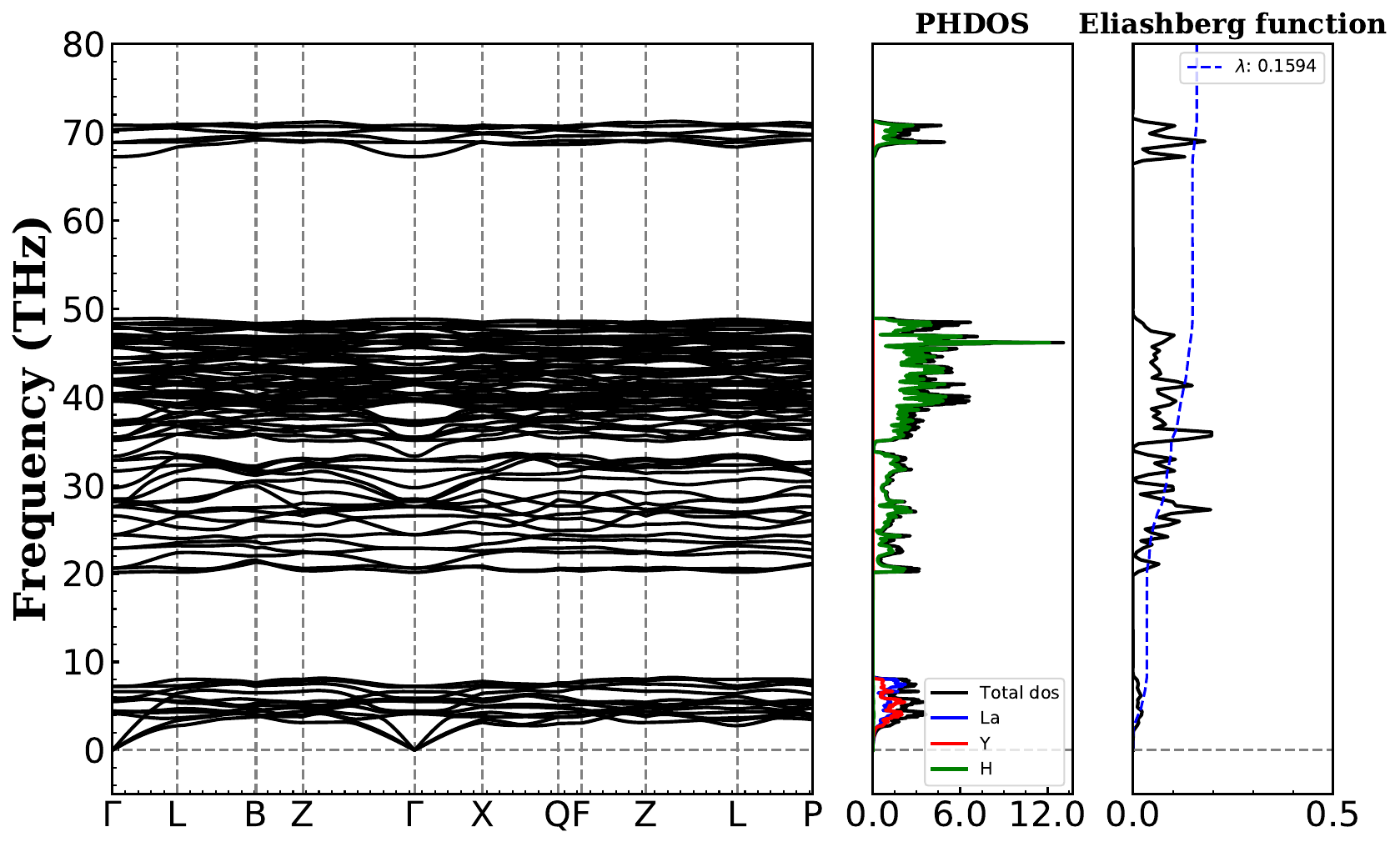}
    \caption{
      Phonon dispersions and projected phonon
      densities of states 
      for $R\bar{3}c$-LaYH$_{12}$ clathrate at  100 GPa.
    }
    \label{fig.phonon_LaYH12_167}
  \end{center}
\end{figure*}

\begin{figure*}
  \begin{center}
    \includegraphics[width=0.5\linewidth]{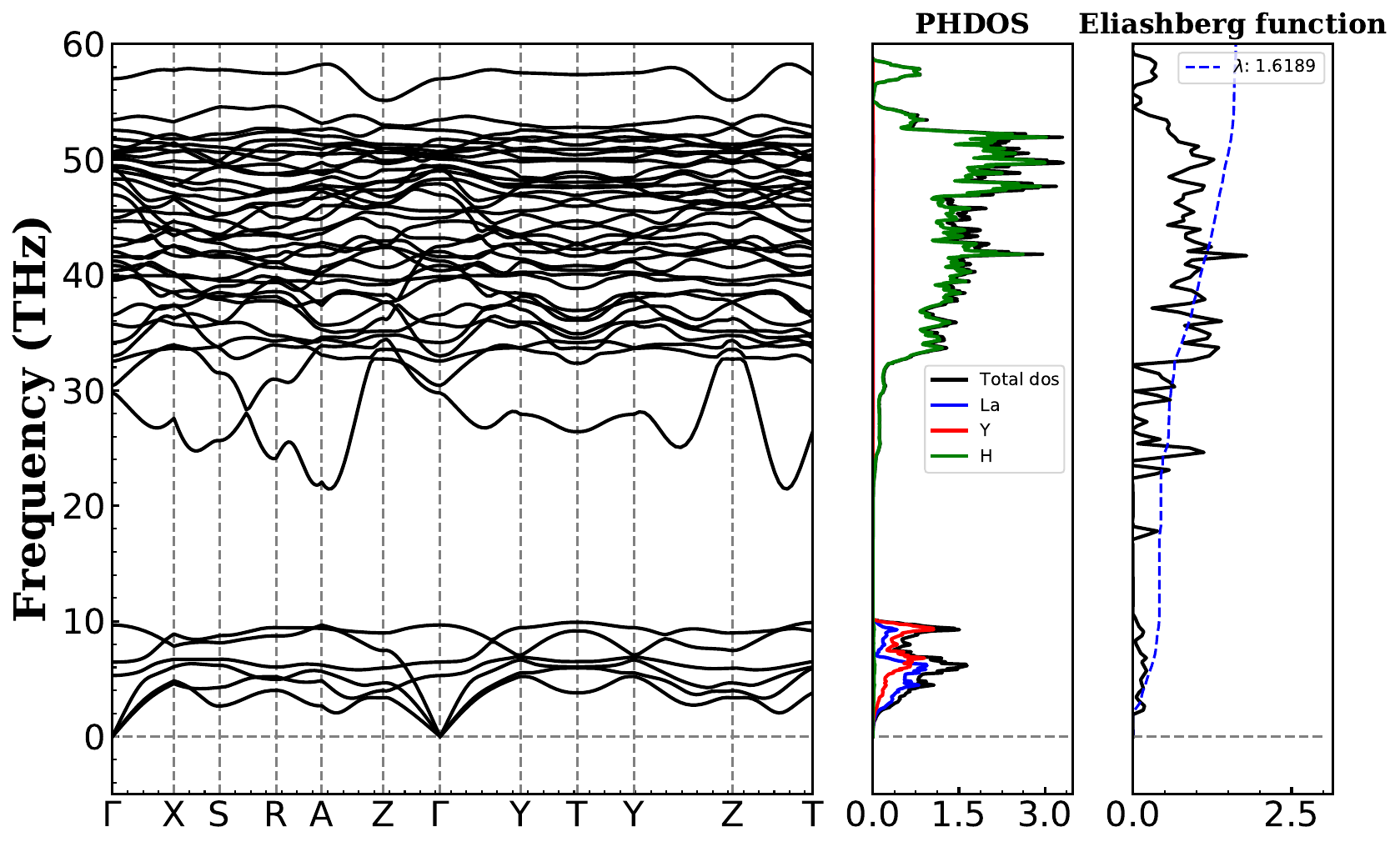}
    \caption{
      Phonon dispersions, projected phonon
      densities of states, and Eliashberg spectral function 
      for $Cmmm$-LaYH$_{12}$ clathrate at 250~GPa.
    }
    \label{fig.phonon_LaYH12_65_250GPa}
  \end{center}
\end{figure*}

\begin{figure*}
  \begin{center}
    \includegraphics[width=0.5\linewidth]{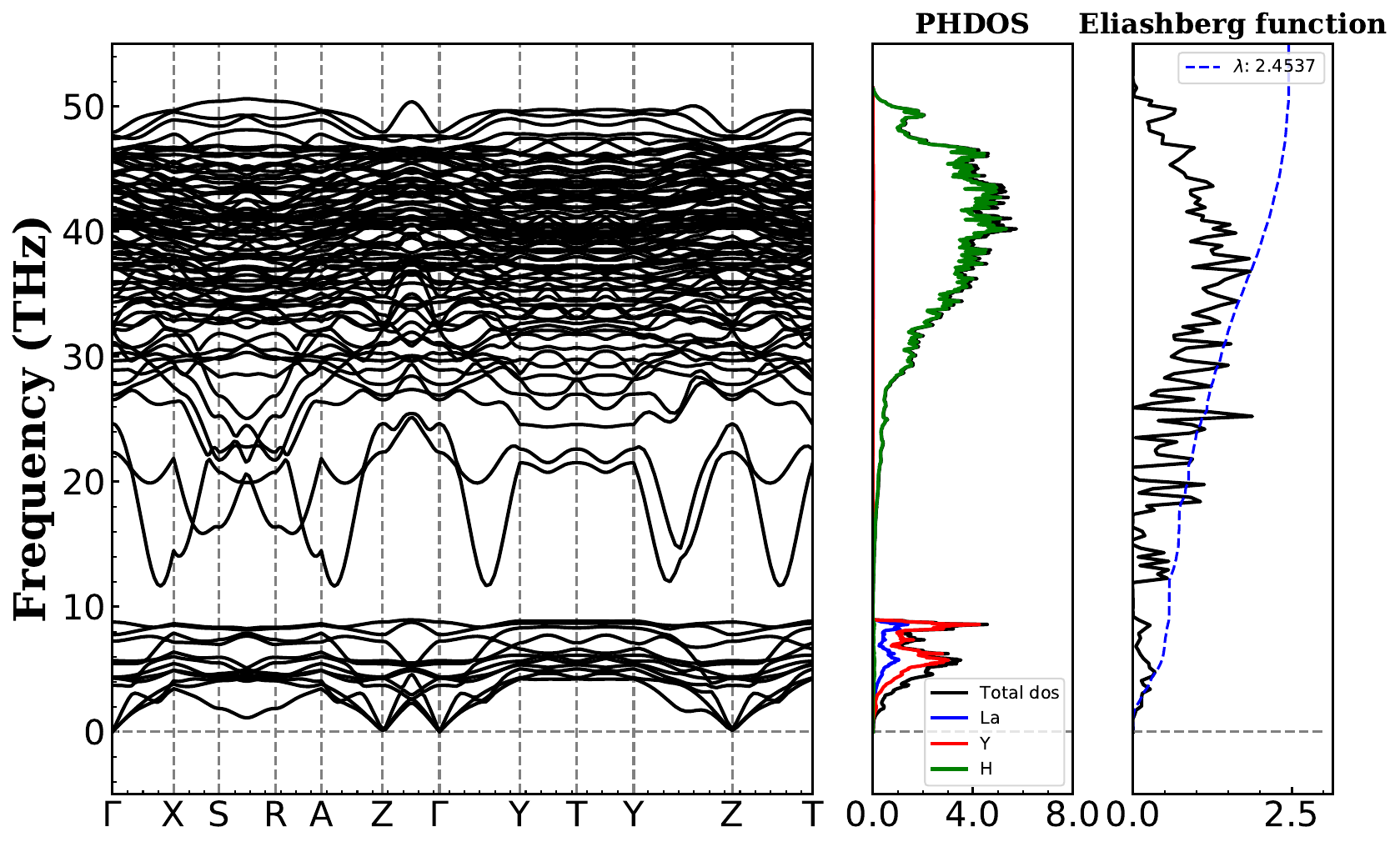}
    \caption{
      Phonon dispersions, projected phonon
      densities of states, and Eliashberg spectral function 
      for $Cmmm$-LaY$_{3}$H$_{24}$ clathrate at 180~GPa.
    }
    \label{fig.phonon_LaY3H24_65}
  \end{center}
\end{figure*}

\begin{figure*}
  \begin{center}
    \includegraphics[width=0.7\linewidth]{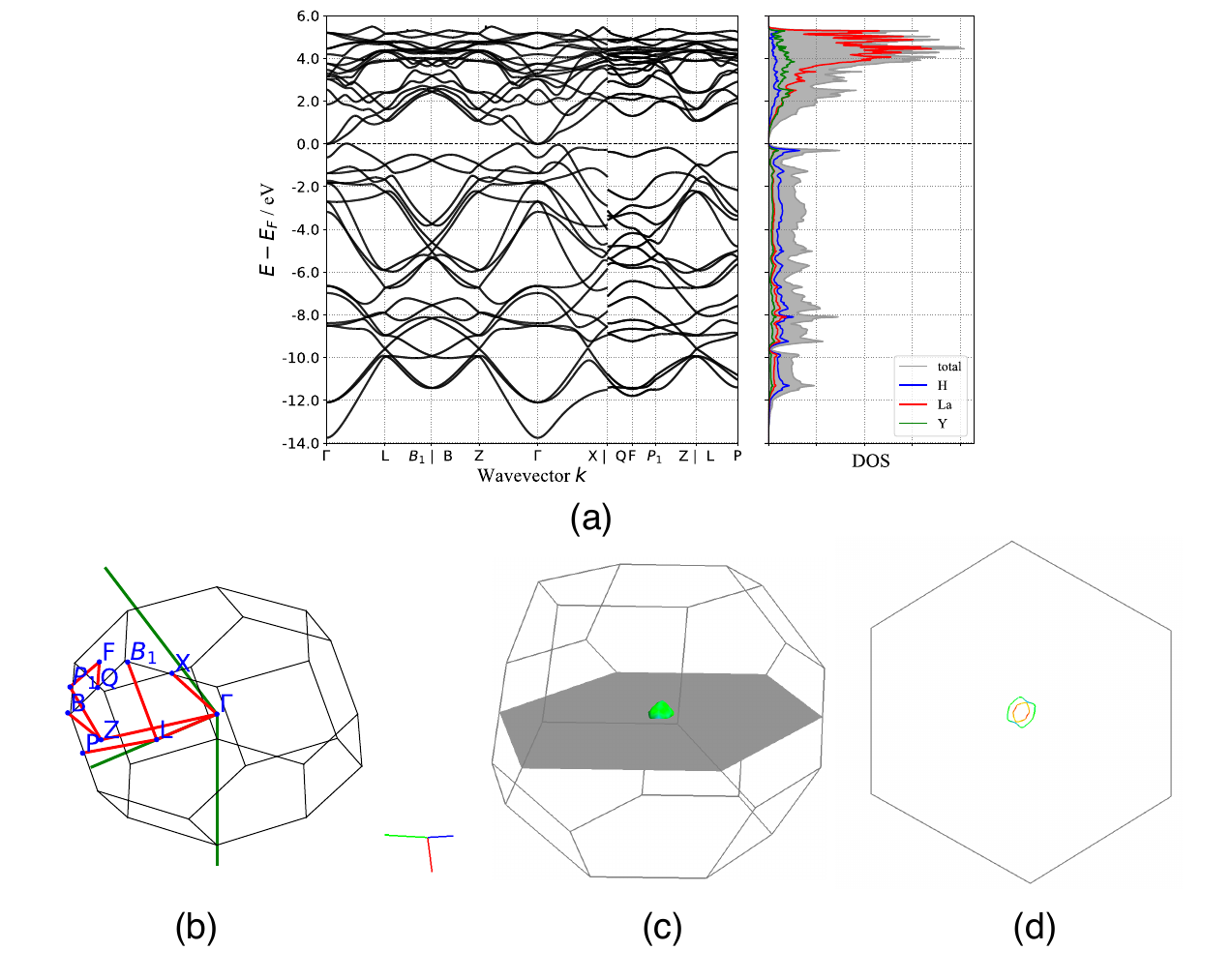}
    \caption{
      Band structure, DOS, and Fermi surface of $R\bar{3}c$-LaYH$_{12}$.
    }
    \label{fig.LaYH12_band_dos_167}
  \end{center}
\end{figure*}

\begin{figure*}
  \begin{center}
    \includegraphics[width=0.7\linewidth]{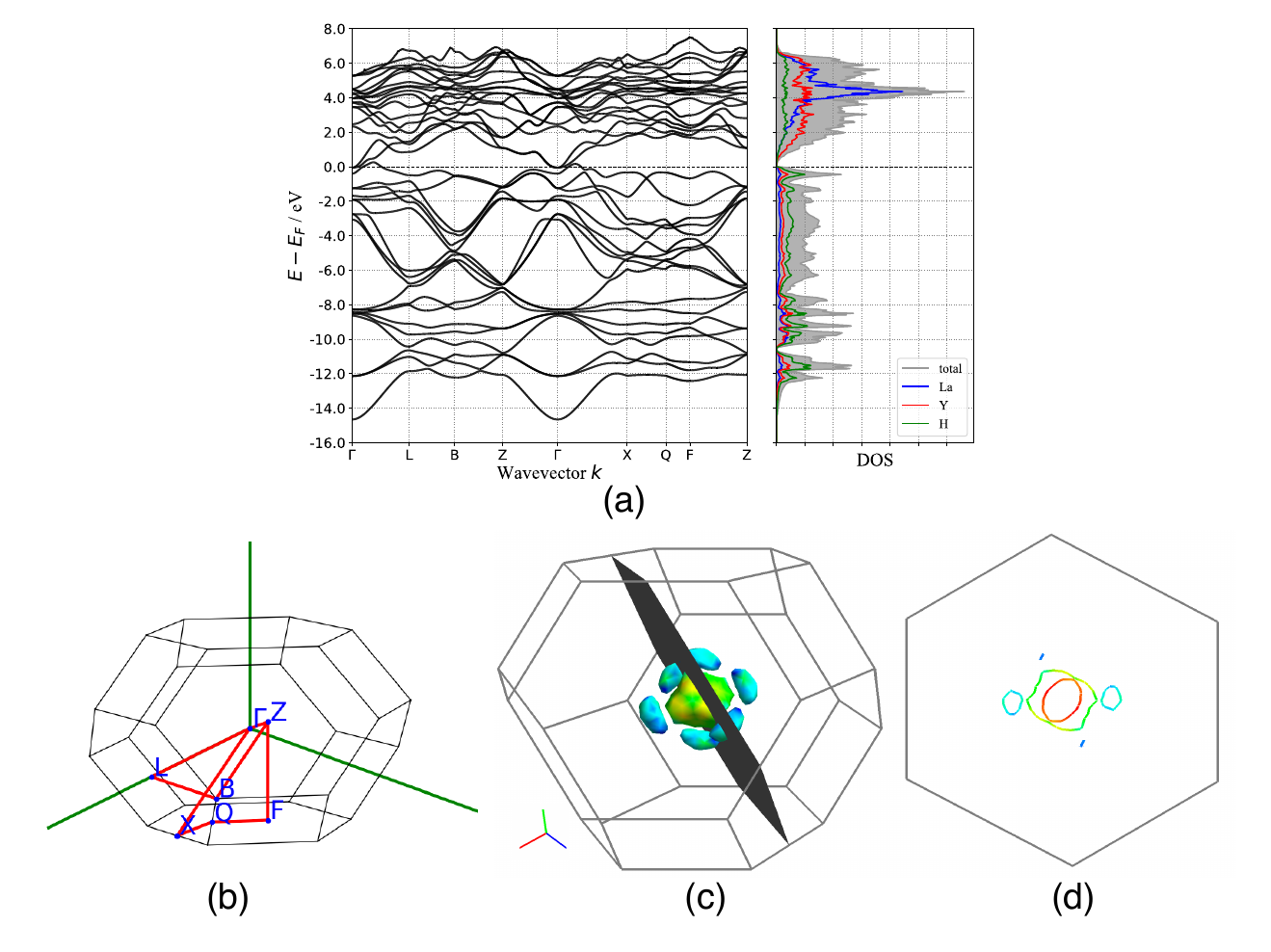}
    \caption{
      Band structure, DOS, and Fermi surface of $R\bar{3}$-LaY$_{3}$H$_{24}$.
    }
    \label{fig.LaY3H24_band_dos_148}
  \end{center}
\end{figure*}

\begin{figure*}
  \begin{center}
    \includegraphics[width=0.7\linewidth]{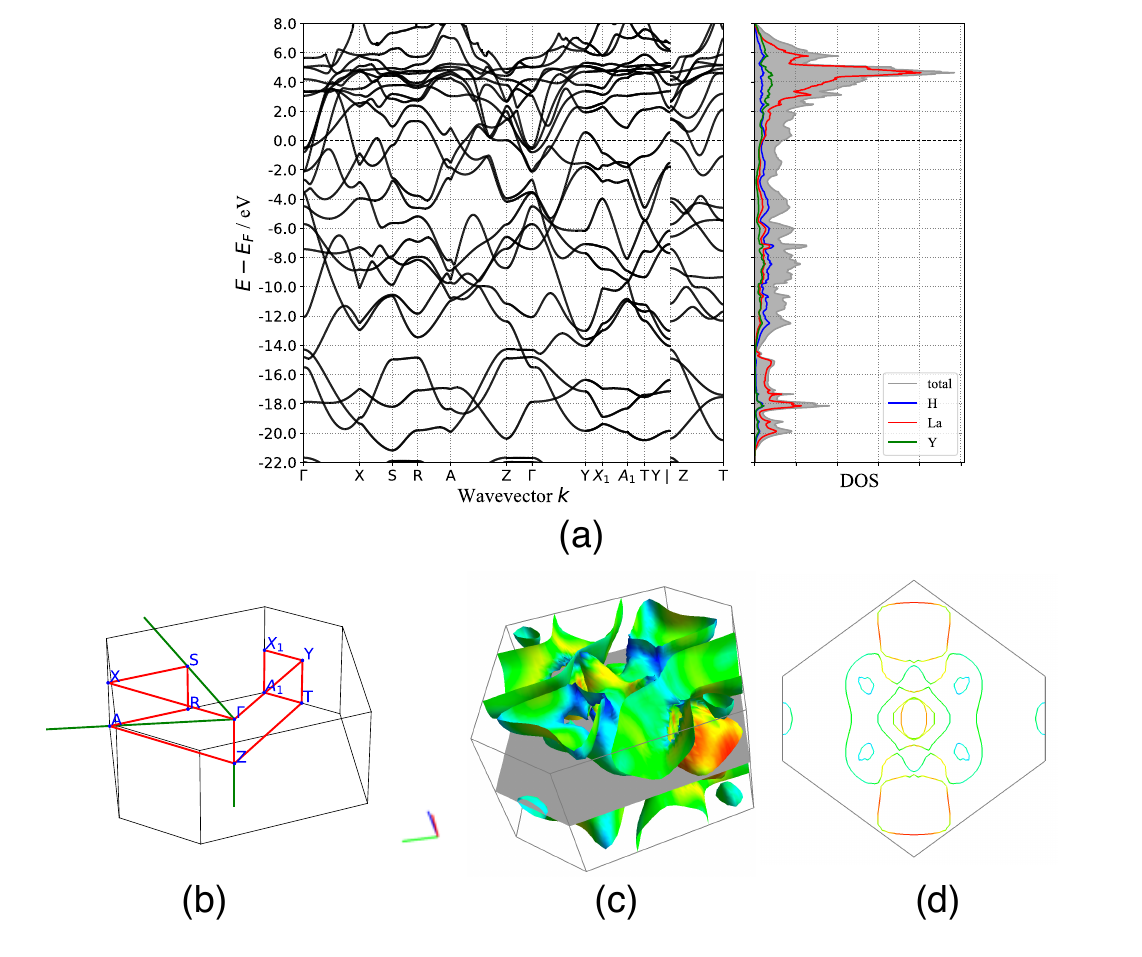}
    \caption{
      Band structure, DOS, and Fermi surface 
      of $Cmmm$-LaYH$_{12}$
    }
    \label{fig.LaYH12_band_dos_65}
  \end{center}
\end{figure*}

\begin{figure*}
  \begin{center}
    \includegraphics[width=0.7\linewidth]{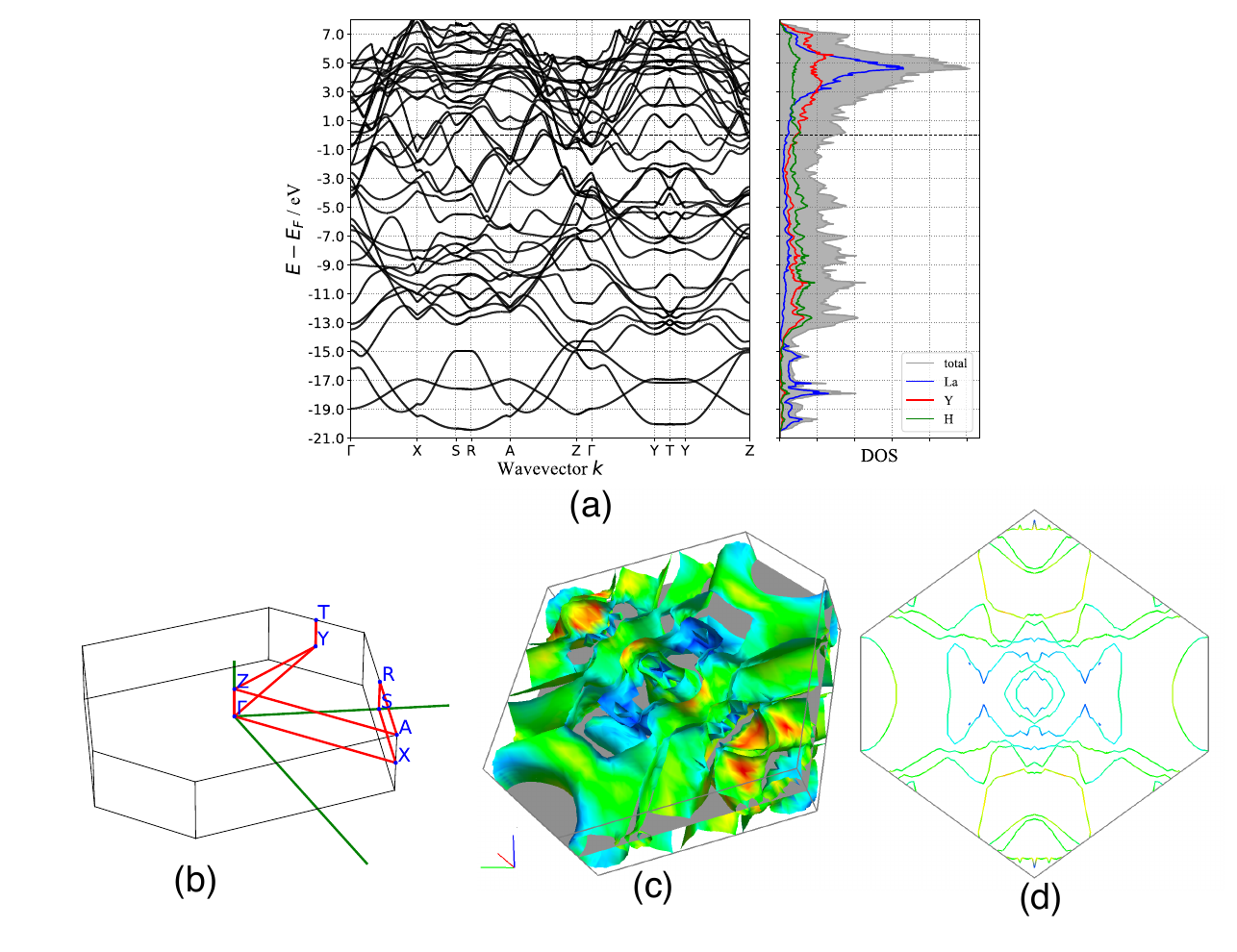}
    \caption{
      Band structure, DOS, and Fermi surface 
      of $Cmmm$-LaY$_{3}$H$_{24}$
    }
    \label{fig.LaY3H24_band_dos_65}
  \end{center}
\end{figure*}

\begin{figure*}
  \begin{center}
    \includegraphics[width=0.6\linewidth]{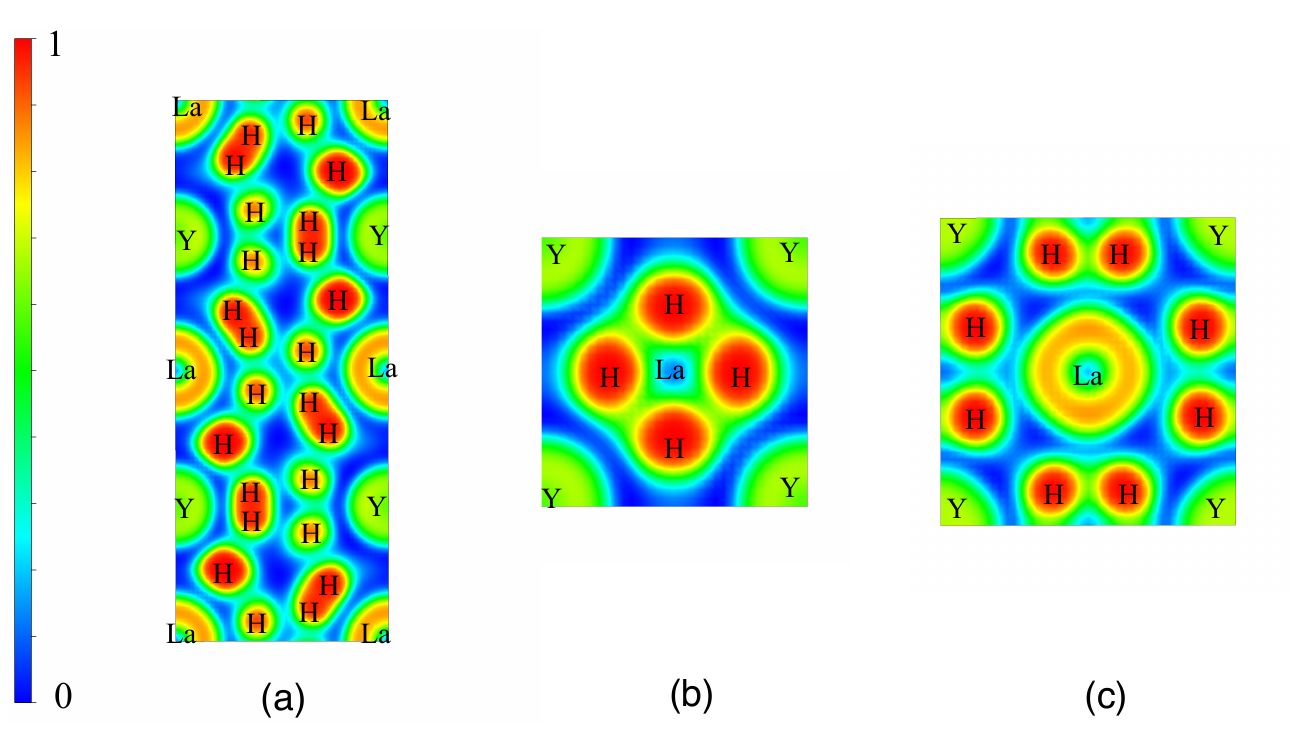}
    \caption{
      The calculated ELF of
      (a) $R\bar{3}c$-LaYH$_{12}$,
      (b) $Pm\bar{3}m$-LaYH$_{12}$, and
      (c) $Cmmm$-LaYH$_{12}$ in the (100) plane.
    }
    \label{fig.LaYH12_elf}
  \end{center}
\end{figure*}

\begin{figure*}
  \begin{center}
    \includegraphics[width=0.5\linewidth]{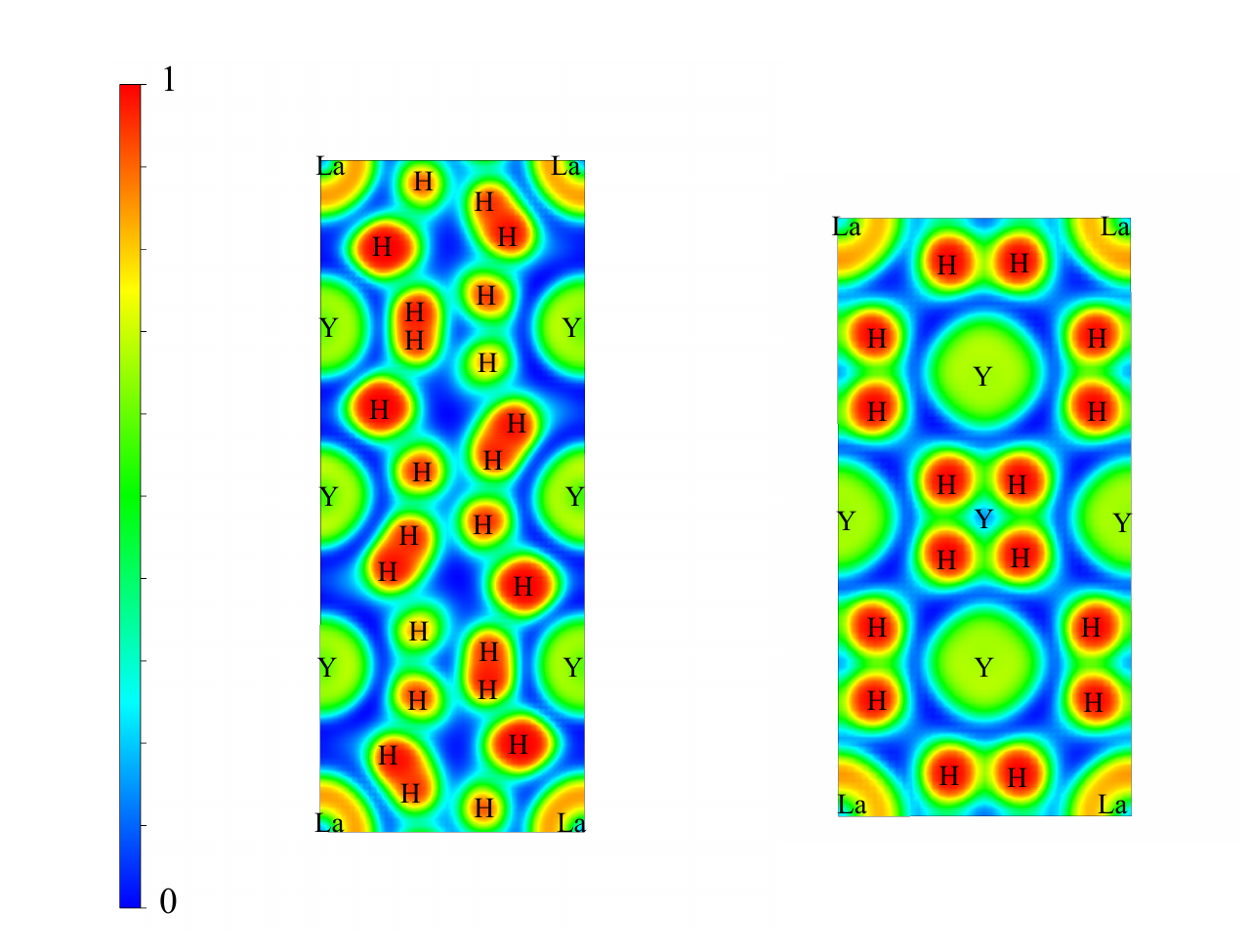}
    \caption{
      The calculated ELF of (a) $R\bar{3}$-LaY$_{3}$H$_{24}$
      and (b) $Cmmm$-LaY$_{3}$H$_{24}$
      in the (100) plane.
    }
    \label{fig.LaY3H24_elf}
  \end{center}
\end{figure*}

\begin{figure*}
  \begin{center}
    \includegraphics[width=0.7\linewidth]{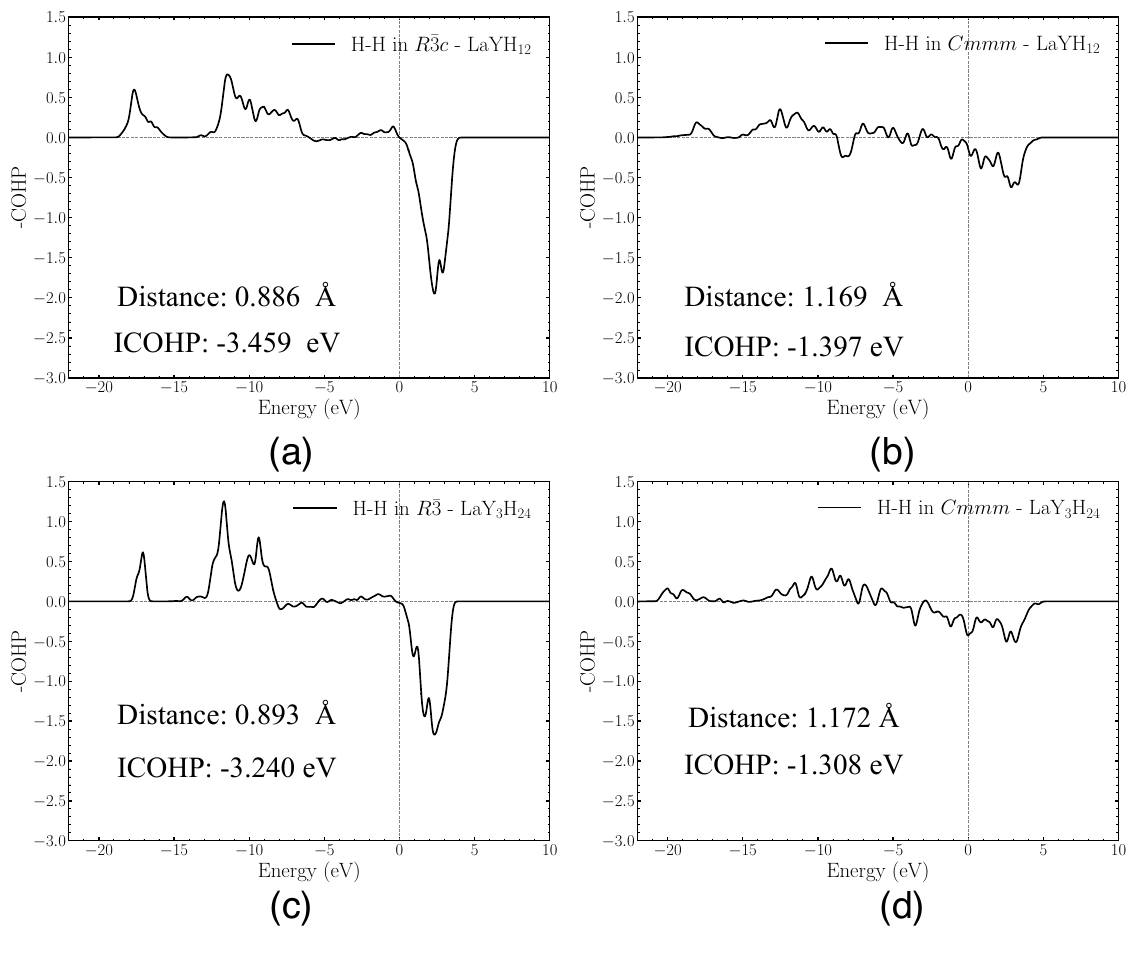}
    \caption{
      COHP for pairs of H-H in (a) $R\bar{3}c$-LaYH$_{12}$ at 100~GPa,
      (b) $Cmmm$-LaYH$_{12}$ at 200~GPa,
      (c) $R\bar{3}$-LaY$_{3}$H$_{24}$ at 100~GPa, 
      and
      (d) $Cmmm$-LaY$_{3}$H$_{24}$ at 180~GPa.
    }
    \label{fig.cohp}
  \end{center}
\end{figure*}

For each compound reported in the main text, 
we provide the detailed geometries 
[Table~\ref{table.crytal_struc})], 
Phonon dispersions and 
Eliashberg spectral functions 
[Figs.~\ref{fig.phonon_LaYH12_167}-\ref{fig.phonon_LaY3H24_65}], 
and electronic band structures and Fermi surfaces 
[Figs.~\ref{fig.LaYH12_band_dos_167}-\ref{fig.LaY3H24_band_dos_65}]. 
To show the bonding characteristics of the hydrogen atoms 
in each compound, 
we also provide the results 
of the electron localization function (ELF) 
[Figs.~\ref{fig.LaYH12_elf} and \ref{fig.LaY3H24_elf}], 
and the crystal orbital Hamilton population 
(COHP) [Fig.~\ref{fig.cohp}].
\begin{table*}
 \begin{center}
   \caption{
     Crystal structures of
     LaYH$_{12}$ and LaY$_{3}$H$_{24}$ 
     predicted at each pressure~($P$).
     Lattice parameters ($a$, $b$, and $c$)
     are given in units of $\AA$.
   }
     \label{table.crytal_struc}
\begin{tabular}{c|c|c|r|crrr}
  & &\multirow{2}{*}{$P$~(GPa)} & \multirow{2}{*}{Lattice parameters}
  & \multicolumn{4}{l}{Atomic coordinates (fractional)} 
\\ \cline{5-8}
&& &  & Atoms & $x$  & $y$  & $z$
  \\
\hline
LaYH$_{12}$    & $R\bar{3}c$ & 100 & \begin{tabular}[c]{@{}l@{}}
  $a = b = 5.389$ \\
  $c = 13.674$  \\
  $\alpha = \beta = 90^{\circ}$
  \\ $\gamma = 120^{\circ}$
  \end{tabular}
  & \begin{tabular}[c]{@{}r@{}}
   La(6$b$)  \\
  Y(6$a$)  \\
  H(36$f$)  \\
  H(36$f$)
  \end{tabular}
  & \begin{tabular}[c]{@{}r@{}}
0.00000 \\
0.00000 \\
0.00181 \\
0.00492
  \end{tabular}
  & \begin{tabular}[c]{@{}r@{}}
0.00000 \\
0.00000 \\
0.23464 \\
0.27691
  \end{tabular}
  & \begin{tabular}[c]{@{}r@{}}
0.00000 \\
0.25000 \\
0.36866 \\
0.60940 
\end{tabular} \\
\hline
LaYH$_{12}$
& $Cmmm$
& 200
& \begin{tabular}[c]{@{}r@{}}
$a =  3.617$ \\
$b = 4.956$ \\
$c = 5.156$
\\
$\alpha = \beta = \gamma = 90^{\circ}$
\\
\end{tabular}
& \begin{tabular}[c]{@{}r@{}}
La(2$c$) \\
Y(2$a$) \\
H(8$n$) \\
H(8$n$) \\
H(8$m$)
\end{tabular}
& \begin{tabular}[c]{@{}r@{}}
0.00000 \\
0.00000 \\
0.00000 \\
0.00000 \\
0.25000
\end{tabular}
& \begin{tabular}[c]{@{}r@{}}
0.50000 \\
0.00000 \\
0.11589 \\
0.38205 \\
0.25000
\end{tabular}
& \begin{tabular}[c]{@{}r@{}}
0.50000 \\
0.00000 \\
0.35519 \\
0.11708 \\
0.23317
\end{tabular}
\\
\hline
LaY$_{3}$H$_{24}$
& $R\bar{3}$
& 100
& \begin{tabular}[c]{@{}r@{}}
$a = b =  5.324$ \\
$c = 13.569$
\\
$\alpha = \beta = 90^{\circ}$\\
$\gamma = 120^{\circ}$

\\
\end{tabular}
& \begin{tabular}[c]{@{}r@{}}
La(3$a$) \\
Y(6$c$) \\
Y(3$b$) \\
H(18$f$) \\
H(18$f$) \\
H(18$f$)\\
H(18$f$)
\end{tabular}
& \begin{tabular}[c]{@{}r@{}}
0.00000 \\
0.00000 \\
0.00000  \\
0.00154 \\
0.00245\\
0.00465\\
0.01123
\end{tabular}
& \begin{tabular}[c]{@{}r@{}}
0.00000 \\
0.00000 \\
0.00000 \\
0.26668 \\
0.23293\\
0.75887\\
0.72593
\end{tabular}
& \begin{tabular}[c]{@{}r@{}}
0.00000 \\
0.24782 \\
0.50000 \\
0.60934 \\
0.36741\\
0.87000\\
0.11041
\end{tabular}
\\
\hline
LaY$_{3}$H$_{24}$
& $Cmmm$
& 180
& \begin{tabular}[c]{@{}r@{}}
$a = 3.621$ \\
$b = 5.013$\\
$c = 10.239$
\\
$\alpha = \beta =  \gamma = 90^{\circ}$\\

\\
\end{tabular}
& \begin{tabular}[c]{@{}r@{}}
La(2$a$) \\
Y(4$l$) \\
Y(2$d$) \\
H(8$n$) \\
H(8$n$) \\
H(8$n$)\\
H(8$n$)\\
H(8$m$)\\
H(8$m$)
\end{tabular}
& \begin{tabular}[c]{@{}r@{}}
0.00000 \\
0.00000 \\
0.00000 \\
0.00000 \\
0.00000\\
0.00000\\
0.00000\\
0.25000\\
0.25000
\end{tabular}
& \begin{tabular}[c]{@{}r@{}}
0.00000\\
0.50000\\
0.00000\\
0.11988 \\
0.12176\\
0.37658\\
0.38314\\
0.25000\\
0.25000
\end{tabular}
& \begin{tabular}[c]{@{}r@{}}
0.00000 \\
0.25606 \\
0.50000 \\
0.19483 \\
0.31540\\
0.43798\\
 0.07100\\
0.13624\\
0.37798 
\end{tabular}
\\
\hline
\end{tabular}
 \end{center}
\end{table*}

\end{document}